\documentclass[iop]{emulateapj}
\usepackage{color}
\usepackage{natbib}
\slugcomment{}

\shorttitle{NGC~1851: merger of two clusters?}
\shortauthors{Carretta et al.}

\begin{document}
\title{Abundances for a large sample of red giants in NGC~1851: hints for a
merger of two clusters?\altaffilmark{1}}

\author{E. Carretta\altaffilmark{2},
R.G. Gratton\altaffilmark{3},
S. Lucatello\altaffilmark{3,4,5},
A. Bragaglia\altaffilmark{2},
G. Catanzaro\altaffilmark{6},
F. Leone\altaffilmark{7},
Y. Momany\altaffilmark{3,8},
V. D'Orazi\altaffilmark{3},
S. Cassisi\altaffilmark{9},
F. D'Antona\altaffilmark{10},
S. Ortolani\altaffilmark{11}
}

\altaffiltext{1}{Based on data collected at the ESO telescopes under 
programme 083.D-0208}
\altaffiltext{2}{INAF, Osservatorio Astronomico di Bologna, via Ranzani 1,
       40127,  Bologna,  Italy. eugenio.carretta@oabo.inaf.it,
       angela.bragaglia@oabo.inaf.it}
\altaffiltext{3}{INAF, Osservatorio Astronomico di Padova, vicolo
       dell'Osservatorio 5, 35122 Padova,  Italy. raffaele.gratton@oapd.inaf.it
       sara.lucatello@oapd.inaf.it, valentina.dorazi@oapd.inaf.it}
\altaffiltext{4}{Excellence Cluster Universe, Technische Universit\"at M\"unchen, 
Boltzmannstr. 2, D-85748, Garching, Germany}
\altaffiltext{5}{INAF-Osservatorio Astrofisico di Catania, Via S.Sofia 78, I-95123 
Catania, Italy}
\altaffiltext{6}{Max Planck Institute for Astrophysics, Postfach 1317, 85741 Garching, Germany}
\altaffiltext{7}{Dipartimento di Fisica e Astronomia, Universit\`a di Catania, Via S.Sofia 78, I-95123 
Catania, Italy}
\altaffiltext{8}{European Southern Observatory, Alonso de Cordova 3107, Vitacura, 
Santiago, Chile}
\altaffiltext{9}{INAF, Osservatorio Astronomico di Collurania, via M.Maggini, 
64100,  Teramo, Italy}
\altaffiltext{10}{INAF, Osservatorio Astronomico di Roma, via Frascati 33, 
00040, Monteporzio Catone (Roma), Italy}
\altaffiltext{11}{Dipartimento di Astronomia, Universit\`a di Padova, Vicolo dell'Osservatorio 2,
35122 Padova, Italy}

\begin{abstract}

We present the abundance analysis of a sample of more than 120 red giants 
in the globular cluster (GC) NGC~1851, based on FLAMES spectra. We find a 
small but detectable metallicity spread.  This spread is compatible
with the presence of two different groups of stars with a metallicity
difference of 0.06-0.08 dex, in agreement with earlier photometric studies. 
If stars are divided into these two groups according to their metallicity, both 
components show a Na-O anticorrelation (signature of a genuine GC nature)
of moderate extension. The metal-poor stars are more concentrated than the metal-rich
ones. We tentatively propose the hypothesis that NGC~1851 formed from a merger of 
two individual GCs with a slightly different Fe and $\alpha-$element 
content, and possibly an age difference up to 1 Gyr. This is supported also by 
number ratios of stars on the split subgiant and on the bimodal horizontal branches. 
The distribution of n-capture process elements in the two components also supports 
the idea that the enrichment must have occurred in each of the structures separately, 
and not as a continuum of events in a single GC. The most probable explanation 
is that the proto-clusters formed into a (now dissolved) dwarf galaxy and later merged 
to produce the present GC.

\end{abstract}


\section{Introduction}

The idea that massive Galactic globular clusters (GCs) may be formed within
dwarf galaxies is not new (e.g., \citealt{bek03}). Recently, theoretical
considerations and new observations  demonstrated that massive GCs
like $\omega$ Cen or M~54 probably had their origin in dwarf spheroidal
galaxies (dSphs), which are currently either lost or still surrounding the cluster
(\citealt{bok09,bel08,car10a}). This idea was extended by
\cite{car10b} to practically $all$ GCs, sketching a scenario which unifies the
view of GCs and dSphs. The ancestral progenitors of both kinds
of systems started as cosmological fragments, but the
evolution of dSphs proceeded undisturbed in near isolation from the
distant main Galaxy. On the contrary, strong interaction with the Galactic main
body triggered a chain of events in the so called $precursor$ of current GCs,
whose final products are the systems we are seeing (see \citealt{car10b} for a
detailed description). The same mechanism may also work within a single dSph,
as in the case of Fornax or Sagittarius. 

This scenario, although not entirely new (e.g., \citealt{sea78}), may 
explain several characteristics of GCs, including the multiple
stellar generations,  found in $all$ objects investigated so far
(\citealt{car09a,car09b} and \citealt{gra04} for a review).
It is currently well assessed that the second-generation stars (presently
constituting the bulk of the cluster population) should have formed from
the ejecta of only a fraction of the first-generation (primordial) stars (e.g.,
\citealt{gra01,pra06}). To account for the present chemical composition, a precursor
baryonic mass  about 20-50 times
larger than the current mass of the GCs is required \citep{car10b}. 
However, the proposed scenario is still qualitative; for
instance, it is not obvious that only $one$ GC should be the final
output of an individual precursor. Examples of binary and multiple clusters
are frequently observed in the Large Magellanic Cloud (e.g.,
\citealt{die02}). 

To clarify these issues, we added NGC~1851 to our ongoing FLAMES 
survey studying the Na-O anticorrelation in GCs \citep{car06}.
NGC~1851 has a bimodal horizontal branch (HB) and several other 
peculiarities. The color-magnitude diagram (CMD) shows a double subgiant 
branch (SGB; \citealt{mil08}). Two distinct red 
giant branch (RGB) sequences were discovered by \cite{lee09a} with Ca$uvby$ 
photometry, and confirmed by \cite{han09} using broad-band filters. In 
\cite{car10c} we  challenged the hypothesis advanced by \cite{lee09b} of Ca 
variations in GCs, ascribed to a possible pollution by core-collapse SN~II. 
However, the case of NGC~1851 was left open, since 
we had not yet adequate data.

A photometric approach alone is probably not enough to understand the complex
nature of multiple populations in GCs.
The most striking photometric feature in NGC~1851, the split SGB, has been
explained in terms either of an age/metallicity/helium effect or a different 
total CNO content between the two populations (e.g., \citealt{cas08,ven09}). 
A mix of two or more factors cannot be excluded. Furthermore, the
spatial
distribution of these sub-populations is still controversial 
(\citealt{zoc09,mil09}).  Only a precise chemical tagging allows the accurate
separation of different stellar generations, and helps providing a first
relative ranking in age between them.  The only existing abundance analysis from
high resolution spectroscopy in this GC is the one by \cite{yon08}, on eight RGB
stars. That analysis (complemented by \citealt{yon09}) showed a number of
additional interesting features: a possible small metallicity spread,
correlations between abundances of p-capture elements and elements produced in
$s-$processes, and possibly a variable value of the total C+N+O sum.  However,
these data are only available for a frustrating small number of stars.

In this Letter we partially fill this gap, presenting the results on the
chemical composition of  more than  120 red giants. We think that our results
support the hypothesis that NGC~1851  formed starting from two clusters (likely
born within a single dSph), an idea already circulating in the literature for
GCs with composite CMDs, in particular for NGC~1851 (e.g.
\citealt{van96,cat97}). These two clusters have slightly different metallicity,
and later underwent a  merger, leaving however detectable traces of that past
event.

\section{Data and analysis}

The FLAMES spectra (GIRAFFE and UVES) of 124 RGB members of NGC~1851  were
obtained in April, August, and September 2009. The stellar parameters were
determined using the same techniques  described, e.g., in \cite{car10a}. 
A full description of the analysis will be presented elsewhere
(Carretta et al., in preparation); here we only show abundances
of some interesting species: Fe, Na, O, Ca, a few $s-$process elements, such
as Ba, La, Ce, and the $r-$process element Eu. 
Details of our abundance analysis trace as closely as possible the homogeneous
procedures adopted for other GCs (\citealt{car09a,car09b} and references
therein)\footnote{For instance, the temperature is derived from a relation
between $T_{\rm eff}$ (from $V-K$ and the \citealt{alo99} calibration) and $K$
magnitudes, much more reliably measured than colors, leading to very small
internal errors in the atmospheric parameters, hence in the derived
abundances.}

\section{Metallicity spread in NGC~1851}

Our first result {\it is that there is a small but real spread in metallicity}
in NGC~1851. We find an average  [Fe/H]{\sc i} $=-1.179 \pm 0.019$ 
($\sigma=0.067$ dex) from 13 stars with UVES spectra. The analysis of
the GIRAFFE spectra yields [Fe/H]{\sc i} $=-1.158 \pm 0.005$ 
($\sigma=0.051$ dex, 121 stars), 
after correcting the equivalent widths to the
system defined by the higher resolution UVES spectra (\citealt{car07a}). 
The observed dispersions in [Fe/H]{\sc i} for UVES spectra are statistically 
significant when compared to internal errors in [Fe/H], which are estimated to
be 0.017 dex (internal errors for GIRAFFE spectra, 0.060 dex, are too large
to be used in this context)\footnote{For a detailed discussion of the larger
scatter found  among brighter and cooler stars see  \cite{car09c}. {\it The
observed spreads we found are only lower limits}, due to the  way we derive the
final temperatures, using a $mean$\ relation between  T$_{\rm eff}(V-K)$ and the
$K$ magnitude along the RGB, which tends to decrease real abundance spreads. 
Finally, the reddening is very low \citep[$E(B-V)$=0.02,][]{har96}  and not
differential.}.

\begin{figure} 
\includegraphics[width=9cm]{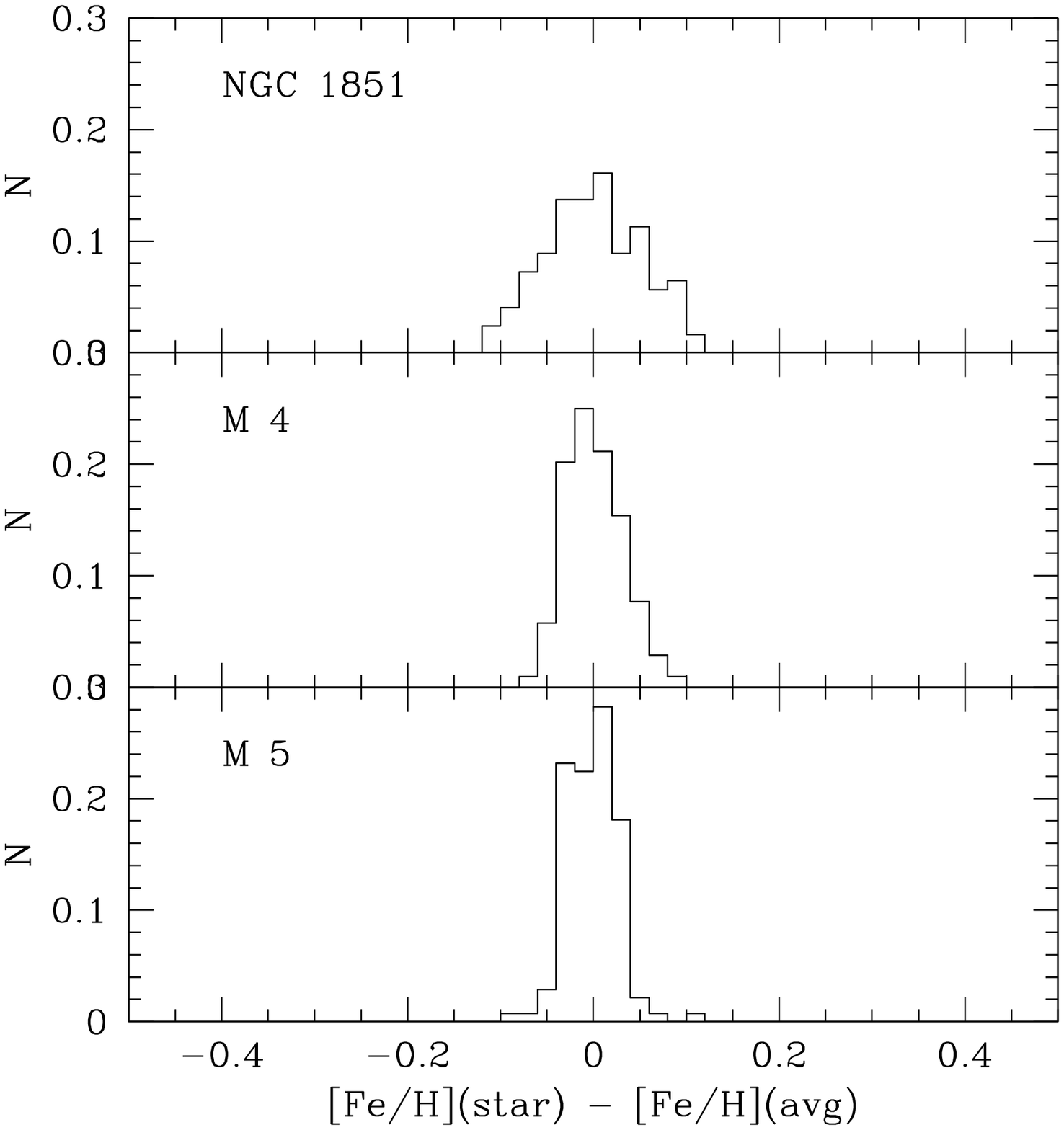}
\includegraphics[width=9cm]{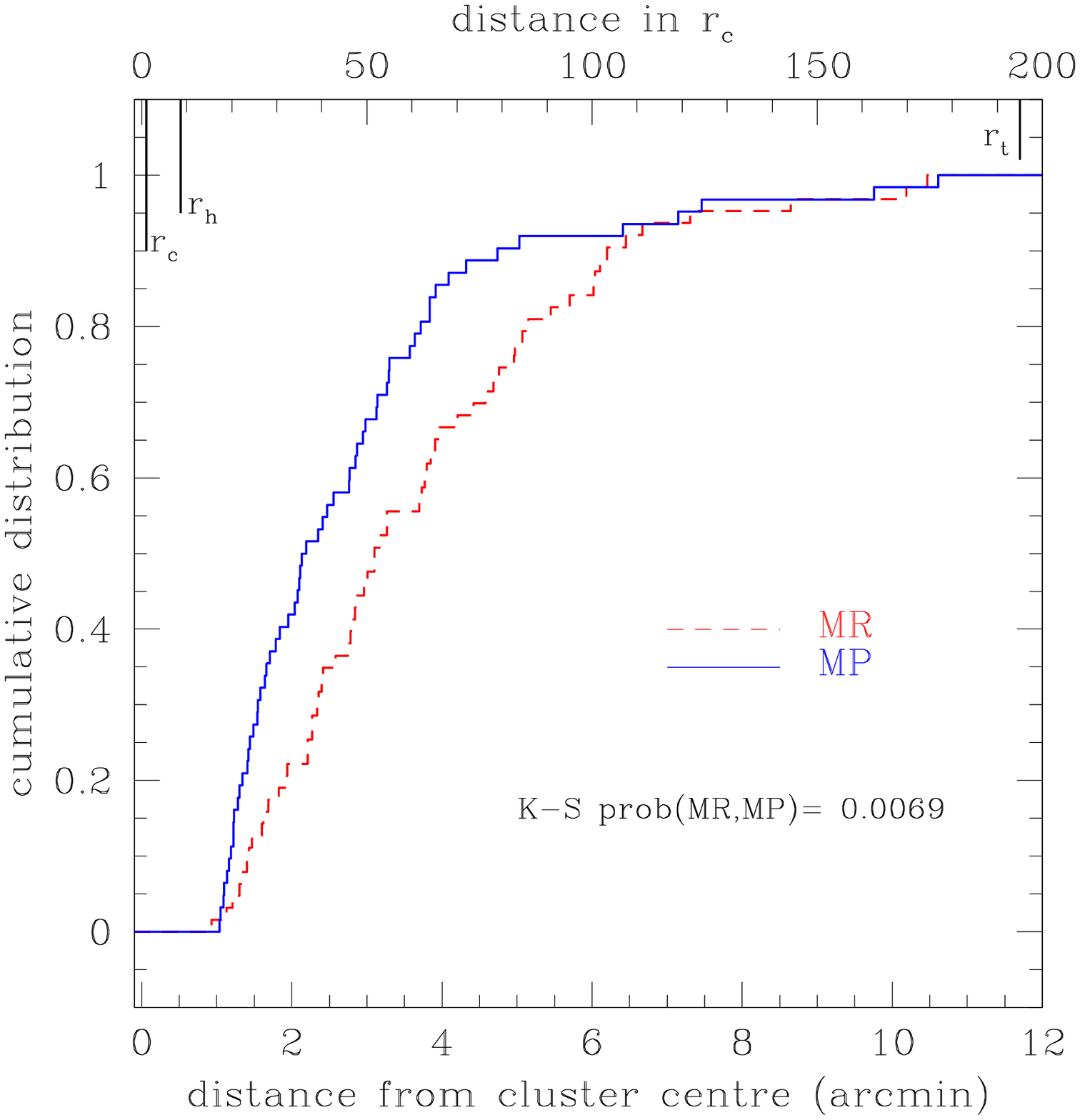}
\caption{Left: MDF of NGC~1851
compared to those in M~4 and M~5 (\citealt{car09a}). All MDFs are
normalized to the cluster average [Fe/H] value. Right: cumulative distributions
of the radial distances for the MR and MP components in NGC~1851.  A
Kolmogorov-Smirnov test returns a negligible probability ($\sim0.007$) that the
two come from the same distribution.  Our stars are all from $\sim 2$ 
half mass radii to the tidal radius.}
\label{f:fig1}
\end{figure}

There are additional ways  to show that the spread in [Fe/H] is real. First,
\cite{yon08} also found some range in [Fe/H], although they
did not expand on this due to the small size of their sample. Second,  in
Fig.~\ref{f:fig1} we compare the metallicity distribution function (MDF) in
NGC~1851 with the MDFs of M~4 and M~5, GCs of similar
metallicity  analyzed through an identical procedure (\citealt{car09c}). 
The metallicity spread in
NGC~1851 is clearly larger.  The  MDF of M~4 would reproduce the one of NGC~1851
if we split it into two equal components with a difference in metallicity of
0.06-0.08 dex; the same is obtained for M~5. Finally, we also find a spread in
[Ca/H], similar to the one in [Fe/H], confirming photometric results by
\cite{lee09a} and \cite{han09}.

\begin{figure} 
\includegraphics[width=9cm]{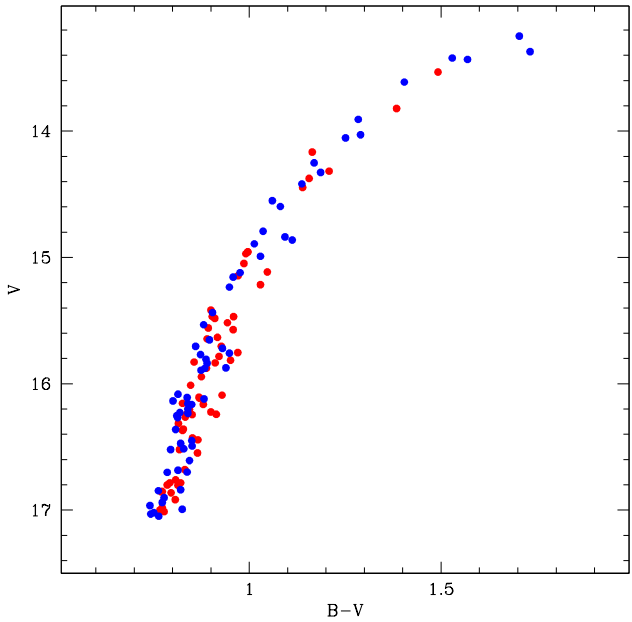}
\caption{ CMD $V,B-V$ indicating MP and MR stars in blue and red, respectively. }
\label{f:fig2}
\end{figure}

For the sake of the following discussion we arbitrarily assume that NGC~1851 is
made of a metal-poor  (MP) and a metal-rich (MR) component. Since there is no
obvious gap between these components, we simply assume that stars with
[Fe/H] larger than the average -1.16 dex belong to the MR component, the others
to the MP one.  Fig.~\ref{f:fig2} shows the RGB accordingly divided. If we fit
the RGB with a line and compute the residuals in color between the individual
stars and the line, the MR stars have average $B-V$  color redder by
0.02$\pm$0.008 mag than the MP ones, in very good agreement with what is
expected (\citealt{gra10}) from the difference in metallicity. 

One of our most striking results is shown in the right panel of Fig.1: 
{\it these two metallicity components have a clearly different spatial
distribution}, with the MP one being more concentrated
than the MR component.  All our stars are in the external regions, at
more than 2 half-mass radii (Fig.~\ref{f:fig1}), where the relaxation time is very
long so that mass segregation between the two components cannot be responsible,
since the mass difference implied by the different metallicity and age (see Sect.
6) is about 0.04 M$_\odot$. Our new result, statistically very robust, is fully
independent from any claimed radial distribution of SGB stars.

\section{The Na-O anticorrelation in NGC~1851}

\begin{figure} 
\includegraphics[scale=0.80]{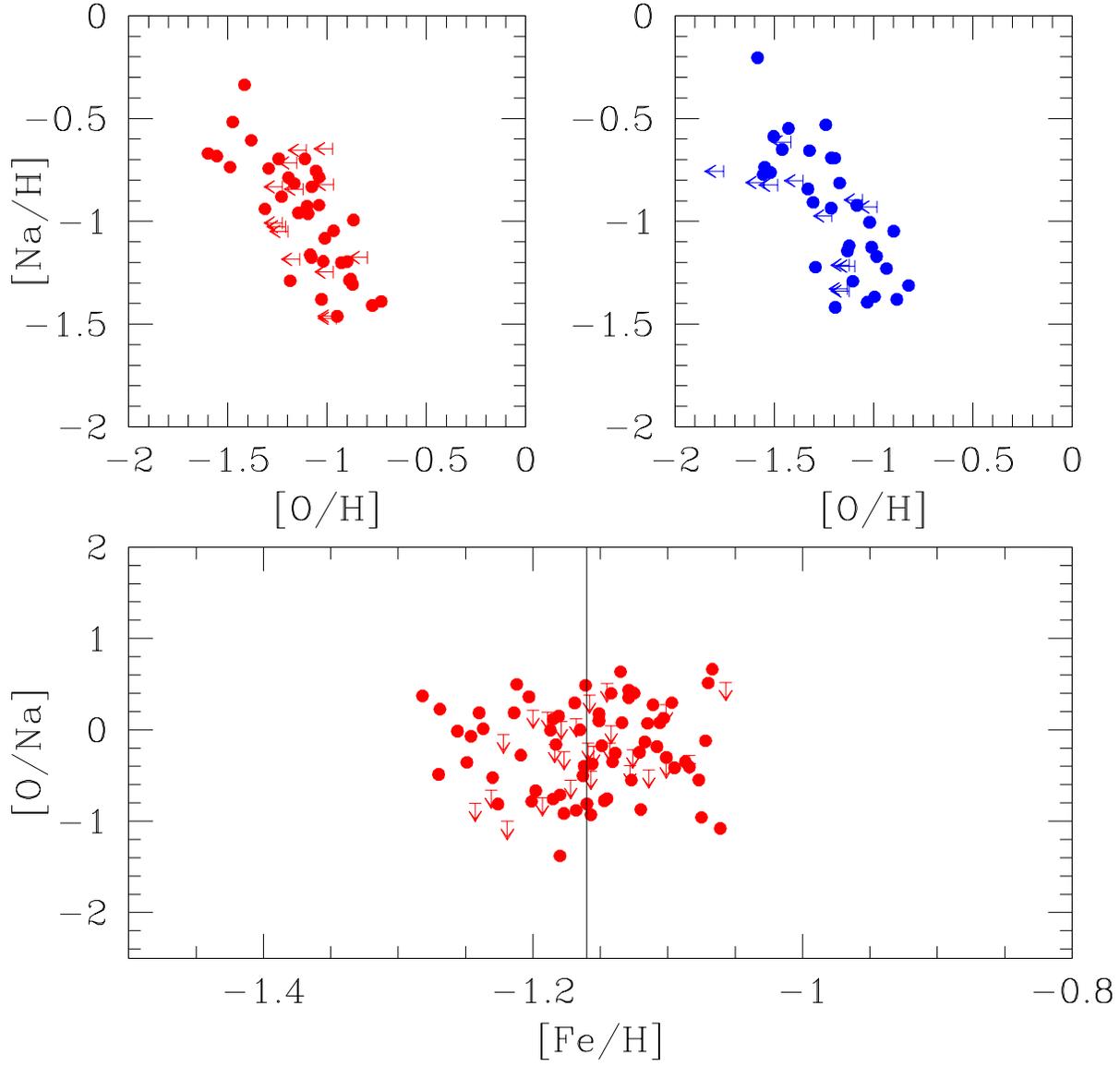}
\caption{Na-O anticorrelation for the MR (upper left panel) and the MP
component (upper right panel). The ratios [Na/H] and [O/H] are used
to remove the dependence on metallicity of the Na abundances of P stars. Bottom
panel: the [O/Na] ratio as a function of metallicity. Upper limits in O are
indicated by arrows.}
\label{f:fig3}
\end{figure}

We find the familiar Na-O anticorrelation also in NGC~1851. Interestingly,
this signature remains visible even considering the MR and MP
populations separately. In Fig.~\ref{f:fig3}  we show the Na-O anticorrelation for both
components, together with the run of the [O/Na] ratios as a function of the
metallicity. We use [Na/H] and [O/H] to account for the
well known metallicity-dependence of Na as a function of the metal-abundance for
stars of the primordial (P) component (see the definition of primordial,
intermediate I, and extreme E populations in \citealt{car09a}).

The Na-O anticorrelation is slightly different for the MR and MP components, 
but in both cases the extension is quite modest, more similar to that in 
M~4 than in NGC~2808. This suggests only a modest spread in He in NGC~1851 
because a high Y fraction seems to be associated only to very long tails of very O-poor 
stars,  not present in this GC. Our finding from a chemical approach
confirms the claims by \cite{sal08}, based on the absence of a tilt along the HB
and the lack of a splitting in the MS.

In addition, we observe a slight change of the mean value of O and Na abundances 
at the level of the bump on the RGB. We already showed 
\citep{car07b,bra10} that this variation is expected from theoretical
models, which predict a change in the bump luminosity with He 
content \citep{sal06}, i.e. with 
elements involved in p-capture reactions. 
The presence of a mix of first and second  generation stars results in a
concentration of Na-poor/He-poor stars just before  the bump accompanied by an
accumulation of Na-rich/He-rich stars just 
above the bump level. This accounts for the observed abundance changes  at the
bump without resurrecting the internal mixing scenario for O and Na  (e.g.
\citealt{lee10}) and then overcoming the unpalatable requirement of basic 
stellar structure differences between field and GC stars. For NGC~1851,  where
we hypothesize two distinct clusters (see below), we expect a further smearing 
of the bump in the RGB luminosity function (Carretta et al., in prep).

\section{The location of stars on the RGB in NGC~1851}

\begin{figure} 
\centering
\includegraphics[bb=25 167 283 713, scale=0.80]{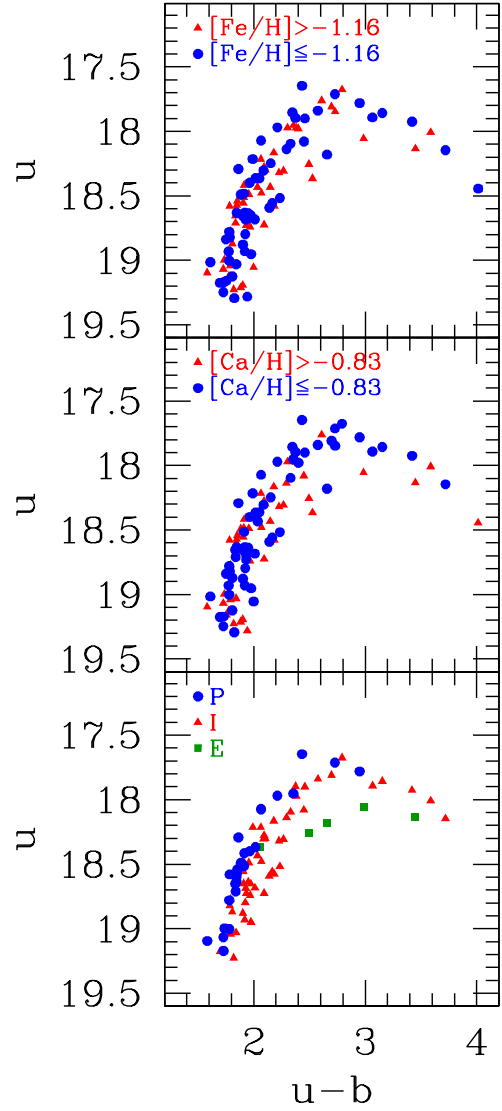}
\caption{$u,u-b$ CMDs for stars analyzed in NGC~1851.
Top: stars are separated at the average metallicity 
[Fe/H]=-1.16 
in MR (red filled triangles) and MP (blue filled circles).   Middle: the
separation is made at the average [Ca/H]=-0.83 value (Ca-rich: red filled
triangles, Ca-poor blue filled circles). Bottom: stars are separated in P (blue
filled circles), I (red filled triangles) and E (green filled squares) 
according to their Na,O abundances.}
\label{f:fig4}
\end{figure}

In Fig.~\ref{f:fig4}, we use the Str\"omgren $u,u-b$ CMD to test where stars of
different components and populations
are located on the RGB. 
This plane is optimally suited to separate first and second
generation stars, probably because of N (enhanced in O-depleted,
second generation stars) via the formation of NH, CN and their relevance on the $u-b$
(or the  Johnson $U-B$), see \cite{yon008,mar08,car09a}.

Stars of the first generation (P, \citealt{car09a}) in NGC~1851 lie along a narrow
strip to the blue 
of the RGB (Fig. \ref{f:fig4}, bottom panel), as
expected from their unprocessed chemical abundances. On the contrary, the
second generation stars  are spread out to the red, as in NGC~6752
(\citealt{car09a}): the I stars are in the middle and the extreme E
component, with the lowest O abundances, is located at the reddest edge.
This segregation is followed also within each metallicity
component, and it is ``orthogonal" to the separation of MR and MP stars
(Fig.~\ref{f:fig4}, top panel), which are well intermingled across all the RGB in this
color. The same holds if we separate the RGB stars using the average value for Ca
([Ca/H]$=-0.83$): stars with low and high Ca are spread across the entire RGB 
(Fig. ~\ref{f:fig4}, middle panel).

Therefore the spread of Ca  does not track the abundances of
p-capture elements. A K-S test on the cumulative distributions of [Ca/H] for 
stars of the first and second generations in NGC~1851 indicates that they
are indistinguishable. Instead, the Ca abundances  
closely track those of Fe. The cumulative distribution of [Ca/H] values for 
the MR and MP components on the RGB are definitively different. Moreover, also 
the radial distributions of Ca-rich and Ca-poor stars confirm the close 
correspondence with metallicity: the Ca-poor giants are more concentrated,
while Ca-rich stars show a tendency toward more 
external regions.

\section{Is NGC~1851 a relic of a merger of two clusters?}

Is there a comprehensive scenario able to account for all the evidence found here
and in previous works in NGC~1851? In our view, the answer is affirmative if we
consider NGC~1851 as the result of a chain of events that started with two
distinct clusters. Several suggestions of duplicity come from the bimodal
distribution of HB stars, the double SGB, and hints of double sequences on the
RGB. Up to now, the main objection was the absence of a metallicity spread,  
owing to the lack of precise abundances 
for a statistically significant number of
stars. This limitation has finally been overcome by our study.

As a tentative working hypothesis we can think of two different clusters,
born in a much larger system, perhaps a dSph. Being distinct, each one might have
formed with a slightly different metallicity and with a different level of
$\alpha-$elements\footnote{We find that also Mg, Si, and Ti track iron in
the MR and MP components, as Ca does.}. Each object is rightfully a GC, since each 
component show the Na-O anticorrelation, the classical
signature of the processes ending in a GC (\citealt{car10b}). After a while, the
two clusters underwent a merger, likely because both were dragged to the center of the 
dSph by dynamical friction (see \citealt{bel08}) and the result is NGC~1851. 
Finally, the dSph merged with the Milky Way. We think that this is the simplest 
scenario, that with a minimum of hypothesis may account for many observational 
constraints. 

The two MR and MP components do not show any significant difference in
kinematics, the velocity dispersion being the same for both components. A
comprehensive dynamical model would be very welcome, although we do not know
when the merging occurred. We will rely on the observed chemistry.

The observables include: a double SGB, where the faint SGB (fSGB) includes
45\% of the stars and the bright SGB (bSGB) the remaining 55\% (\citealt{mil08}),
and with controversial evidence of different concentration; the MR and MP
components on the RGB, with a clear difference in radial concentration; a
bimodal distribution on the HB, with $\sim 40\%$ of the stars on the BHB
and $\sim 60\%$ on the RHB (\citealt{mil08}); the observed luminosity of HB stars
and the moderate extension of the Na-O anticorrelation in both the MR and MP RGB 
components, which both suggests small He abundance variations.

We may explain these observables in different ways:

(i) A single GC with two populations having a different total CNO abundance
(but a similar He abundance). This may explain the SGB but fails to
reproduce the number ratios on the HB (if the same efficiency for 
mass loss on the RGB is assumed for both sub-populations), because the minor fSGB component 
should be associated with the major RHB one. 

(ii) A single GC with two populations having a different He and total 
CNO abundance. In this scenario there is much more He in CNO-rich than in 
CNO normal stars. The CNO effect dominates on the SGB, and He effect on 
the HB. However, in this scenario the BHB should also be brighter than 
the RHB, which is not observed (e.g. \citealt{sal08}).

(iii) A merging of two GCs differing in age (although this does not exclude
{\it a priori} a difference in the total CNO). In this scenario, one of the GC
(including some 55-60\% of the stars) would be responsible for the bright, younger 
(possibly less concentrated) SGB, of the MR (less concentrated) RGB, and of the 
younger RHB. The other GC (including some 40-45\% of the stars) would be responsible 
of the faint, older (possibly more concentrated) SGB, of the MP (more concentrated)
RGB, and of the older BHB. In this case, the two SGBs can be fitted by isochrones 
differing in age by $\sim 1.5$ Gyr, if the same CNO content is assumed for
both GCs, and the difference in Fe is as derived from the MP and MR
RGB components.

Should the solution (iii) be the right one, the age difference that we need 
($\sim 1-1.5$~Gyr) is not unlikely if the clusters were born in a
dSph (see e.g., the case of the Fornax dSph GCs,  \citealt{buo98, buo99}). 
As a consistency check, we explored Lee's diagram HB-type vs [Fe/H]
(see e.g., Fig. 9 in \citealt{rey01})\
using the average values of metallicity [Fe/H]$=-1.20$ ($\sigma=0.03$ dex) and
[Fe/H]$=-1.12$ ($\sigma=0.03$ dex) for the MP and MP components, respectively. If we
tentatively adopt for the two components the HB type of NGC~288 (0.98) and
NGC~362 (-0.87), the diagram  shows that the MR
component may easily be about 1-1.5 Gyr younger then the MP one.

\begin{figure} 
\includegraphics{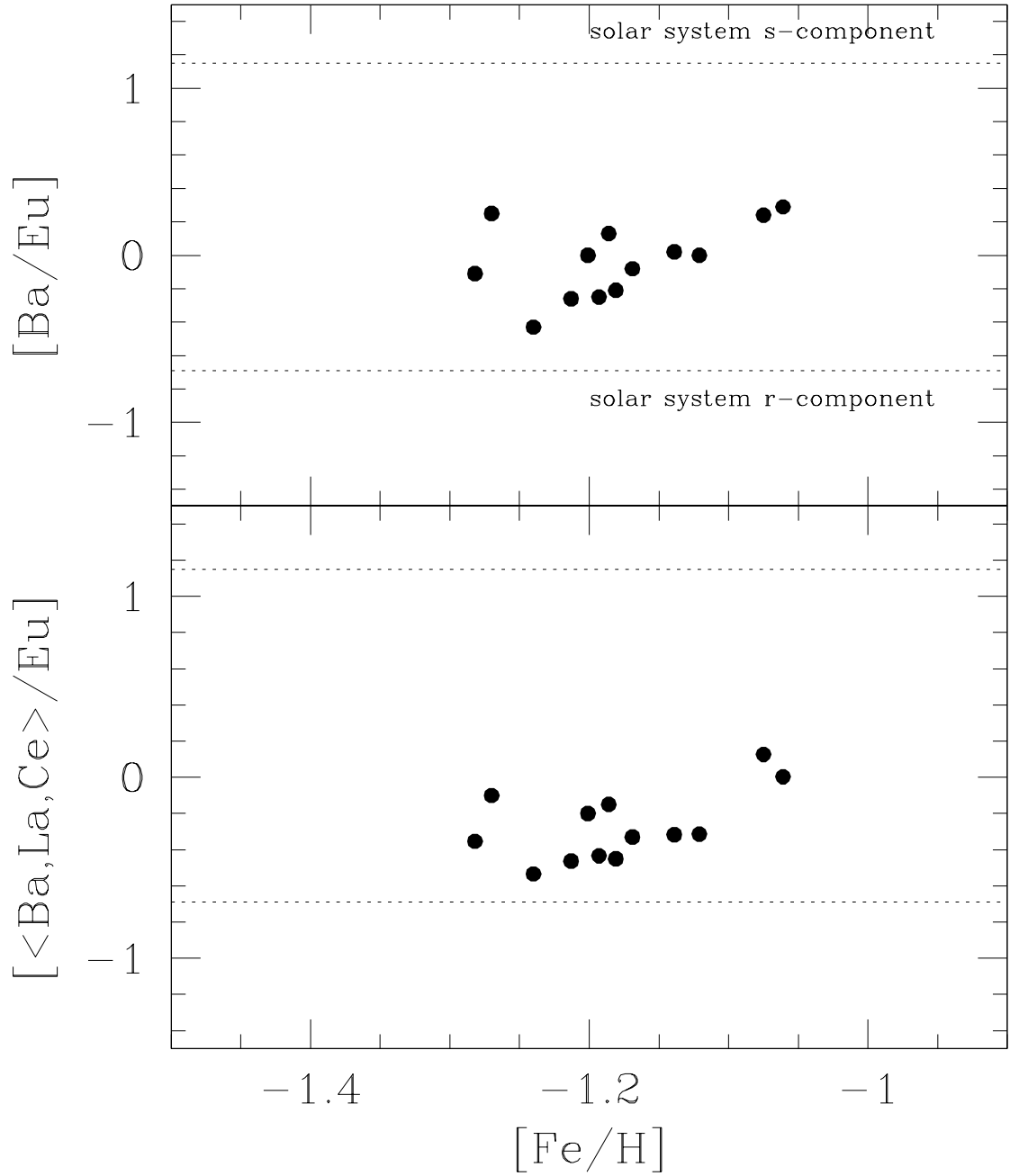}
\caption{Ratio of [$s/r$] process elements for stars with UVES
spectra, as a function of metallicity. Upper panel: [Ba/Eu] ratios. In the 
lower panel, the level of $s-$process elements is represented by the average of
Ba, La and Ce.}
\label{f:fig5}
\end{figure}

An additional constraint comes from the ratio between the abundances of $s-$ 
and $r-$process elements. Fig. ~\ref{f:fig5} shows the [Ba/Eu] ratio as a function of 
metallicity for the 13 stars with UVES spectra. This ratio is close to that
expected from a pure-$r$ component for MP stars, while it indicates a larger 
contribution by $-s$ process for the MR ones. The trend with [Fe/H] is even
cleaner when using the average from Ba, La, and Ce. It suggests a larger 
contribution of polluters of smaller masses for the MR component, which fits 
well with the proposed age difference.

To conclude, we note that  \cite{carb10} reported the existence of a
distinct metal-poor main sequence around the GCs NGC~1851 and NGC~1904, which
they interpreted as a very low surface brightness stellar system. This is maybe
consistent with the structure identified by \cite{ols09} in the form of a halo
of main-sequence stars surrounding NGC~1851 up to a distance of 250 pc. Both
these observations suggest the existence of a residual structure that might be what
is left by the destruction of the ancestral dwarf where the progenitor of
NGC~1851 originated.

A really clear-cut test for the presence of two distinct GCs is to probe the
Na-O anticorrelation among HB stars. If the merger hypothesis is correct, each
one of the HB populations (coming from individual GCs) should
present the Na-O anticorrelation, as we find in the MR and MP components of the
RGB. On the contrary, for a single proto-cluster scenario, we should expect the
RHB stars be almost all O-rich, with the O-poor stars only confined to the BHB.
Specific proposals of observations aimed to perform this test have been already 
submitted. 

\acknowledgements
Partial funding come from the PRIN MIUR 2007 CRA 1.06.07.05, PRIN INAF 2007 CRA
1.06.10.04,  the DFG cluster of excellence ''Origin and Structure of the
Universe''.

\clearpage

\end{document}